\documentclass[aps,floatfix,preprint,showpacs,preprintnumbers,nofootinbib,superscriptaddress,natbib]{revtex4}

\usepackage{graphicx,float}
\usepackage{epsfig}		
\usepackage{dcolumn}
\usepackage{bm}
\usepackage{subfigure}

\usepackage[all]{xy}
\usepackage{amsmath,upgreek}
\usepackage{amssymb}

\usepackage{pdfpages}
\usepackage{color}
\usepackage{graphicx,epstopdf}
\usepackage[colorlinks,hyperindex]{hyperref}
\usepackage{slashed}
\usepackage{bookmark}
\usepackage{hyperref,color,xcolor}
\hypersetup{hidelinks,backref=true,pagebackref=true,hyperindex=true,colorlinks=true,breaklinks=true,urlcolor= purple}
\hypersetup{%
 colorlinks = true,
 linkcolor = red,
 citecolor = blue
}

\def\0{\mbox{\tiny $0$}}
\def\1{\mbox{\tiny $1$}}
\def\2{\mbox{\tiny $2$}}
\def\3{\mbox{\tiny $3$}}
\def\4{\mbox{\tiny $4$}}
\def\5{\mbox{\tiny $5$}}
\def\6{\mbox{\tiny $6$}}
\def\7{\mbox{\tiny $7$}}
\def\8{\mbox{\tiny $8$}}
\def\9{\mbox{\tiny $9$}}

\def\f14{\mbox{\tiny $\frac{1}{4}$}}

\renewcommand{\baselinestretch}{1.4}

\begin{document}

\title{Phase-space quantum profile of P\"{o}schl-Teller two-level systems}

\author{A. E. Bernardini}
\email{alexeb@ufscar.br}
\affiliation{~Departamento de F\'{\i}sica, Universidade Federal de S\~ao Carlos, PO Box 676, 13565-905, S\~ao Carlos, SP, Brasil.}
\author{R. da Rocha}
\email{roldao.rocha@ufabc.edu.br}
\affiliation{Centro de Matem\'atica, Computa\c c\~ao e Cogni\c c\~ao, Universidade Federal do ABC - UFABC, 09210-580, Santo Andr\'e, Brazil.}
\begin{abstract}
The quantum phase-space dynamics driven by hyperbolic P\"oschl-Teller (PT) potentials is investigated in the context of the Weyl-Wigner quantum mechanics.
The obtained Wigner functions for quantum superpositions of ground and first-excited states exhibit some non-classical and non-linear patterns which are theoretically tested and quantified according to a non-Gaussian continuous variable framework.
It comprises the computation of quantifiers of non-classicality for an anharmonic two-level system where non-Liouvillian features are identified through the phase-space portrait of quantum fluctuations. In particular, the associated non-Gaussian profiles are quantified by measures of {\em kurtosis} and {\em negative entropy}.
As expected from the PT {\em quasi}-harmonic profile, our results suggest that quantum wells can work as an experimental platform that approaches the Gaussian behavior in the investigation of the interplay between classical and quantum scenarios.
Furthermore, it is also verified that the Wigner representation admits the construction of a two-particle bipartite quantum system of continuous variables, $A$ and $B$,
which are shown to be separable under Gaussian and non-Gaussian continuous variable criteria.
\end{abstract}

\date{\today}
\maketitle
 
\section{Introduction}

P\"oschl-Teller (PT) potentials are analytically solvable quantum mechanical potentials for which the one-dimensional Schr\"odinger equation (and some of its extended radial versions) has solutions described by associated Legendre polynomials, $\mathcal{P}^{\mu}_{\lambda}(u)$, when $u$ is identified with either hyperbolic ($\sim \tanh(x)$) or trigonometric ($\sim \tan(x)$) functions of the space coordinate, $x$ \cite{PT}.

On the purely (high energy physics) theoretical front, the intrinsic nonlinearity effects produced by PT potentials \cite{BookA} -- or by some of their modified versions -- have been identified in several scenarios, which include the defect structures of topological (domain walls) and non-topological (bell-shaped lumps) origins \cite{Bas01,Bazeia,AlexRoldao,AlexRoldao2}, the interplay between parity-time symmetry and supersymmetry problems \cite{Pana15}, the construction of topological solutions describing solitary waves throughout Bose-Einstein condensates \cite{Smith}, the mass generation mechanism for exotic spinor fields \cite{Ber12}, and even the quantum problems supported by some curved geometrical background \cite{Das,Jatkar,Witten} which, for instance, involves the warped geometry for braneworld models \cite{Gremm,Barb08,Bertolami}.
From the applied physics perspective, the theoretical background provided by PT potential problems has been demonstrated to be reproducible and manipulable by means of several experimental platforms. For instance, in semiconductor physics, the bound-bound, bound-free and free-free intersubband optical transitions are currently reproduced by quantum wells described by PT potentials designed from realistic ternary alloy based structures \cite{PLAetc}. 
In mesoscale scenarios where semiconductor quantum wells are equally relevant, double quantum dots have been described by various theoretical approaches which include modified numerically solvable PT potentials describing two-electron states and their potential entanglement properties \cite{Rasa12}.
Finally, in the engineering of electronic devices, the thermalization of nonequilibrium spinless electrons in quantum wires \cite{Mick11} can also be mapped into the analytical bound state problem of a quantum particle in a PT potential. 
The applicability of theoretical tools associated to PT problems indeed embraces multidisciplinary sectors in physics, which also includes the two-dimensional transport of quasiparticles in bilayer graphene \cite{Park15}, the modeling optical systems with changing refractive index \cite{Yildirim}, the description of light-matter interaction in superconducting circuits \cite{Sanc19}, and the scattering of matter-wave single soliton and two-soliton molecules by quantum wells \cite{Al11}, namely a large class of problems where the signatures of the quantum behavior in the dynamics of macroscopic objects can be investigated.

Besides their experimental feasibility, the corresponding PT Hamiltonians are analytically solvable in the phase-space, and their corresponding Wigner functions, due to their quasi-Gaussian nature, might be relevant in describing the interplay between microscopic-quantum and macroscopic-classical realities.
In particular, the PT quadratic spectrum (driven by integer quantum numbers) supports a two-level system time-oscillating behavior which reproduces, for instance, the quantum state revivals, and can also be considered in the discussion of theoretical aspects of classical to quantum transitions, non-classicality and quantum correlations. 

In such a context, the theoretical tools specialized to PT scenarios are still incipient.
The correspondence between the Wigner phase-space quasi-probability distribution \cite{Wigner} of a given stationary quantum mechanical wave function to a particular solution of the Liouville equation, with bound states described by semiclassical distributions, have already been considered for scenarios described by the PT ground state solution \cite{Bund,Curtright}.
Otherwise, the Wigner function analytical results \cite{Case} can be extended to the description of quantum superpositions of PT ground and first-excited states, $\vert0\rangle$ and $\vert1\rangle$, as to provide a complete framework for discussing the elementary features involving the above mentioned aspects of non-classicality, which is the main proposal of this work.

The outline of the manuscript is then as follows.
Sec.~II is concerned with the preliminary description of the classical and quantum portraits of PT potentials.
The complete analytical description of Wigner functions for P\"{o}schl-Teller two-level systems involving ground and first-excited states, $\vert0\rangle$ and $\vert1\rangle$, is given in Sec.~III.
The quantum to classical correspondence, as well as the involved relative discrepancies, are discussed in terms of the Wigner flow analysis \cite{Steuernagel3,Liouvillian}, where the quantum distortions are obtained from Wigner currents.
In particular, the two-level system non-Liouvillian behavior is identified through the phase-space pattern of quantum fluctuations.
Through the analytical properties of Wigner functions, in Sec.~IV, the corresponding non-Gaussian profile is quantified by measures of {\em kurtosis} and {\em negative entropy}. Finally, in such a context, two-particle bipartite systems of continuous variables, $A$ and $B$, identified by $\sum_{i,j = 0,1}^{i\neq j}\vert i_{_A}\rangle\otimes\vert j_{_B}\rangle$, give rise to pure states in the phase-space scenario, which are proven to be separable under Gaussian and non-Gaussian criteria.
Our concluding remarks are drawn in Sec.~V.

\section{Classical and quantum portraits}

\begin{figure}[h]
\includegraphics[scale=0.4]{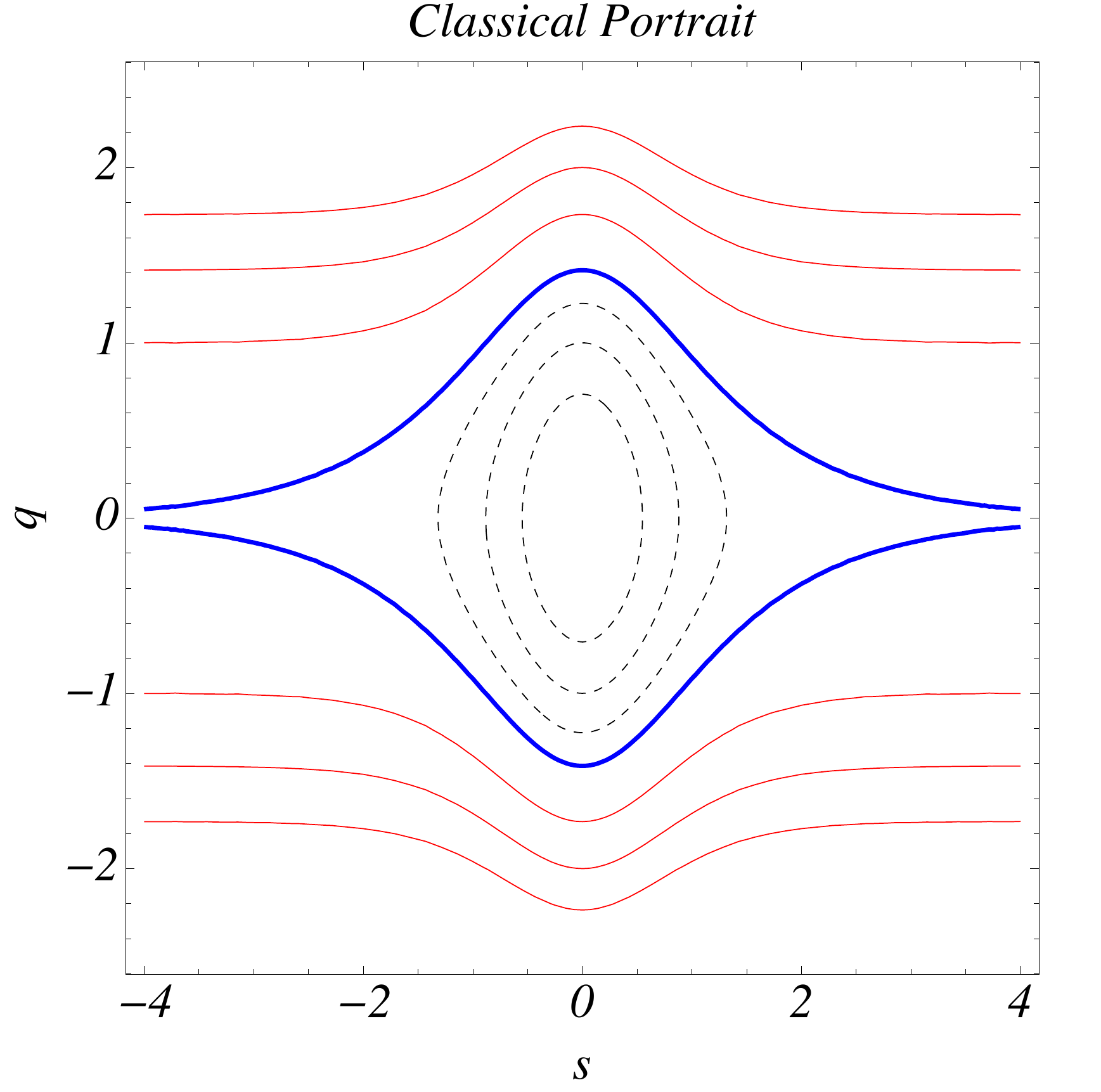}
\renewcommand{\baselinestretch}{.85}
\caption{
(Color online) Classical portrait of P\"{o}schl-Teller Hamiltonians with $\lambda =1$. Phase-space trajectories are for $2 > \mu^2 > 0$ (black dashed lines, for $\mu^2 = 3/2,\,1,\,1/2$), $\mu^2 =0$ (blue thick line) and $\mu^2 < 0$ (red thin lines, for $\mu^2 = -1,\,-2,\,-3$).}
\label{PT}
\end{figure}

The hyperbolic P\"{o}schl-Teller problem for a particle with mass $m$ and momentum $p$ can be introduced through the Hamiltonian
\begin{equation}
H = \frac{p^{\2}}{2m} - \varepsilon \,\lambda(\lambda + 1) \mbox{sech}^2(x/L),
\label{Ham001}
\end{equation}
where $L$ is a width parameter and $\lambda$ defines the minimal potential energy value, $- \varepsilon \,\lambda(\lambda + 1)$, at $x=0$, where the energy dimension is carried by $\varepsilon$. The Hamiltonian can be mapped into a simpler dimensionless version written as
\begin{equation}
H_{\varepsilon} = H \,\varepsilon^{-1} = q^2 - \lambda(\lambda + 1) \,\mbox{sech}^2(s),
\label{Ham001}
\end{equation}
where, to anticipate the connection with quantum mechanical solutions, one has $\varepsilon$ identified by $\hbar^2/2mL$, and the canonical variables given by $s \equiv x/L$ and $q\equiv p L/\hbar$, as to represent the phase-space volume element by $\hbar \,dp\,dx \equiv dq\,ds$.
The classical trajectories are simply given by
\begin{equation}
-\mu^2 = q^2 - \lambda(\lambda + 1)\, \mbox{sech}^2(s),
\label{Ham002}
\end{equation}
where the total energy, $E$, identified by $-\mu^2\,\varepsilon$, defines three classes of solutions valid under constraints given by $\lambda(\lambda + 1) > \mu^2 > 0,\, \mu^2 =0$ and $\mu^2 < 0$, as depicted in the scheme from Fig.~\ref{PT}.

The corresponding Hamiltonian equations of motion result into a non-linear system of equations
\begin{eqnarray}
\label{111}
\dot{s} \equiv \frac{ds}{d\tau} &=& q,\\
\label{222}
\dot{q} \equiv \frac{dq}{d\tau} &=& \lambda(\lambda + 1)\, \tanh(s)\,\mbox{sech}^2(s),
\end{eqnarray}
evolving in terms of a dimensionless time variable, $\tau = 2\varepsilon t/\hbar$.

The analytical solutions for the Hamiltonian system of equations for $\mu^2 < 0$ are obtained as being
\begin{eqnarray}
s(\tau) &=& \pm \mbox{arcsinh}\left[\sqrt{2} \sinh(\sqrt{\lambda(\lambda+1)}\tau)\right],\\
q(\tau) &=& \pm \sqrt{2\lambda(\lambda+1)}\left(\frac{1+\sinh^2(\sqrt{\lambda(\lambda+1)}\tau)}{1+2\sinh^2(\sqrt{\lambda(\lambda+1)}\tau)}\right)^{\frac{1}{2}},
\end{eqnarray}
where the boundary initial values have been given by $s(0) =0$ and $q(0)= \sqrt{2\lambda(\lambda+1)}$ and, by way of simplicity, it has been set $\mu^2 = -\lambda(\lambda+1)$.
Likewise, for $\mu^2 = 0$, with $s(0) =0$ and $q(0)= \sqrt{\ell(\ell+1)}$, one obtains
\begin{eqnarray}
s(\tau) &=& \pm \mbox{arcsinh}\left[\sqrt{\ell(\ell+1)}\tau\right],\\
q(\tau) &=& \pm \left(\frac{\ell(\ell+1)}{1+\ell(\ell+1)\tau}\right)^{\frac{1}{2}}.
\end{eqnarray}
Finally, for $\mu^2 > 0$, with $s(0) =0$ and $q(0)= \sqrt{\ell}$, one has
\begin{eqnarray}
\label{clas1}
s(\tau) &=& \pm \mbox{arcsinh}\left[(1/\sqrt{\ell})\, \sin(\ell\tau)\right],\\
\label{clas2}
q(\tau) &=& \pm \frac{\ell\,\cos(\ell \tau)}{\sqrt{\ell+ \sin^2(\ell \tau)}},
\end{eqnarray}
where, again, by way of simplicity, it has been set $\mu^2 = \ell^2$. In particular, for the classical results where $\mu^2 \geq 0$, the introduction of the arbitrary parameter, $\ell$, is concerned with the comparison with the following quantum results: when quantum state superpositions evolve in time, the role of the parameter $\ell$ will be compared with the role of the parameter $\lambda$ from the Hamiltonian (\ref{Ham002}) since $\ell$ can be described in terms of $\lambda$.

Let us then turn our attention to the quantum mechanical version of the PT Hamiltonian dynamics.
Obviously, Eqs.~\eqref{111}-\eqref{222} reproduce the results for the quantum mechanical Heisenberg equations for quantum operators $\hat{s}$ and $\hat{q}$, since one has identified the commutation correspondence given by $\hbar^{-1}[\hat{x},\, \hat{p}] = [\hat{s},\,\hat{q}] = i$.
In the Schr\"odinger's representation, one has $\hat{s}\rightarrow s$ and $\hat{q} \rightarrow -i\,(d/ds)$ such that the dynamical equation for the wave function $\psi^{\lambda - \mu}_{\lambda}(s)$ reads
\begin{equation}
\left(\frac{d^2}{ds^2} -\mu^2 + \lambda(\lambda + 1) \,\mbox{sech}^2(s)\right)\psi^{\lambda - \mu}_{\lambda}(s) = 0,
\end{equation}
with {\em eigenvalues} $E_{\mu} = -\mu^2 \varepsilon$, where the quantum numbers $\lambda = 1,\, 2,\,\dots$ and $\mu = 1,\,\dots,\, \lambda$ constrain the solutions to 
\begin{equation}
\psi^{\lambda - \mu}_{\lambda}(s) = \mathcal{N}_{\lambda - \mu}(\lambda)\, \mathcal{P}^{\mu}_{\lambda}(\tanh(s)),
\end{equation}
where $\mathcal{N}_{\lambda - \mu}(\lambda)$ are normalization constants, and $\mathcal{P}^{\mu}_{\lambda}$ are the associated Legendre polynomials with an orthonormalization constraint given by
\begin{equation}
\int^{+\infty}_{-\infty}ds\,
\mathcal{P}^{\mu}_{\lambda}(\tanh(s))\mathcal{P}^{\tilde{\mu}}_{\lambda}(\tanh(s)) = 
\frac{\Gamma(\lambda+\mu+1)}{\Gamma(\lambda-\mu+1)\Gamma(\mu+1)}\delta_{\mu\tilde{\mu}},
\end{equation}
where $\Gamma(u)$ is the {\em gamma function}.
Ground and first excited states given by, respectively,
\begin{eqnarray}
\label{p1}
\psi^{0}_{\lambda}(s) &=& \mathcal{A}(\lambda)\, \mbox{sech}^{\lambda}(s),\\
\label{p2}
\psi^{1}_{\lambda}(s) &=& (2(\lambda-1))^{\frac{1}{2}}\,\mathcal{A}(\lambda)\, \sinh(s)\,\mbox{sech}^{\lambda}(s),
\quad\mbox{with}\quad \mathcal{A}(\lambda) = \left(\frac{1}{\sqrt{\pi}}\frac{\Gamma(\lambda+\frac{1}{2})}{\Gamma(\lambda)}\right)^{\frac{1}{2}},
\end{eqnarray}
can compose a two-level system with characteristic frequency given by $\omega = \Delta E_{10}/\hbar = (E_{\lambda-1}-E_{\lambda})/\hbar = (2\lambda - 1)\varepsilon/\hbar$, for which the oscillating phase is given by $(2\lambda - 1)\varepsilon \,t/\hbar = (\lambda - 1/2) \tau$ such that, in the preliminary comparison with the classical dynamics, one may set $\ell = \lambda - 1/2$ into the Eqs.~\eqref{clas1}-\eqref{clas2}.
However, as one will see in the following section, a correspondence between classical and quantum dynamics is much more complex than a simple identification of such time-oscillating behavior.

\section{Wigner function for P\"{o}schl-Teller two-level systems}

The Wigner function is a real-valued quasi-probability distribution defined through the so-called Weyl transform \cite{Wigner,Case}, which exhibits the operational advantage of connecting all the informational content of the state vectors in the phase-space with the interplay between quantum observables and expectation values.
As it has been noticed \cite{Liouvillian,Donoso12,Tunnel,Ballentine}, it provides some tools for the preliminary understanding of the nature of the non-classicality and non-Gaussianity of quantum mechanics, which have been systematically discussed in the literature (see, for instance, \cite{Alba18,Park18} and Refs. therein), and has a particular relevance for the understanding of non-Gaussianity aspects related to quantum correlations \cite{Park17}.

Herein the general features of ground and first-excited stationary states of PT potentials will be discussed in terms of the Wigner formalism, as to provide the analytical tools that give an overall probabilistic interpretation of time-oscillating quantum superpositions, in terms of a Wigner flow description.

The inception of the Wigner function is the definition of the Fourier transform of the off-diagonal coherences of the quantum mechanics density matrix \cite{Wigner,Case} when it is written as $\psi^{*}_{\lambda}(s+y)\,\psi_{\lambda}(s-y)$, for a generic quantum state, $\psi_{\lambda}(s)$, through the expression
\begin{equation}
W_{\lambda}(s,\,q) = \frac{1}{\pi}\int_{-\infty}^{+\infty}dy\, \exp(2i\,q\,y) \,\psi^{*}_{\lambda}(s+y)\,\psi_{\lambda}(s-y),
\label{eqn21}
\end{equation}
where operators $\hat{s}$ and $\hat{q}$ have been converted into commutative numbers, $s$ and $q$, and the dimensionless normalization conditions are expressed by
\begin{equation}
\int_{-\infty}^{+\infty}\hspace{-.3 cm}ds\int_{-\infty}^{+\infty}\hspace{-.3 cm}dq\, W_{\lambda}(s,\,q) = \int_{-\infty}^{+\infty}\hspace{-.3 cm}ds\, \vert \psi_{\lambda}(s)\vert^{2} = 1.
\end{equation}
For a generic quantum state, $\psi^C_{\lambda}(s)$, written as a superposition of ground and first-excited states from Eqs.~\eqref{p1} and \eqref{p2}, $\psi^C_{\lambda}(s) = a_0 \psi^{0}_{\lambda}(s) + a_1\psi^{1}_{\lambda}(s)$, with $|a_0|^2+|a_1|^2=1$, the complete expression for the Wigner function can be put in the form of a sum of elementary contributions given by
\begin{equation}
W^C_{\lambda}(s,\,q) = \sum_{i,j=0,1} a_i^* a_j \,W^{(ij)}_{\lambda}(s,\,q)
\end{equation}
with
\begin{equation}
W^{(ij)}_{\lambda}(s,\,q) = \frac{1}{\pi}\int_{-\infty}^{+\infty}dy\, \exp(2i\,q\,y) \,\psi^{i*}_{\lambda}(s+y)\,\psi^j_{\lambda}(s-y),
\end{equation}
from which one can compute the respective components as
\small\begin{eqnarray}
\label{p1w}
W^{(00)}_{\lambda}(s,\,q) 
&=& 2^{\lambda}\mathcal{A}^{\2}(\lambda)\,\pi^{-1}\,\int_{-\infty}^{+\infty}dy\, \exp(2i\,q\,y)\, \left(\cosh(2s)+\cosh(2y)\right)^{-\lambda}\nonumber\\
&=& 4\frac{\mathcal{A}^{\2}(\lambda)}{\Gamma(\lambda)}\,\left(\frac{(-1)}{\sinh(2s)}\frac{d\,}{ds}\right)^{\lambda -1}
\int_{0}^{+\infty}dy\, \cos(2\,q\,y)\, \left(\cosh(2s)+\cosh(2y)\right)^{-1},\\
\label{p2w}
W^{(11)}_{\lambda}(s,\,q) 
&=& 2^{\lambda}(\lambda -1)\mathcal{A}^{\2}(\lambda)\,\pi^{-1}\,\int_{-\infty}^{+\infty}dy\, \exp(2i\,q\,y)\, \frac{\cosh(2s)-\cosh(2y)}{\left(\cosh(2s)+\cosh(2y)\right)^{\lambda}}\nonumber\\
&=& 2^{\lambda}(\lambda - 1)\mathcal{A}^{\2}(\lambda)\,\left[\frac{2}{\Gamma(\lambda)}\cosh(2s)\left(\frac{(-1)}{\sinh(2s)}\frac{d\,}{ds}\right)^{\lambda -1} - \frac{2}{\Gamma(\lambda-1)}\left(\frac{(-1)}{\sinh(2s)}\frac{d\,}{ds}\right)^{\lambda -2}\right]\nonumber\\
&&\qquad\qquad\qquad\qquad\qquad
\int_{0}^{+\infty}dy\, \cos(2\,q\,y)\, \left(\cosh(2s)+\cosh(2y)\right)^{-1},
\end{eqnarray}\normalsize
which can be manipulated to be written as (cf. the results from Eqs.~(3.983) and (3.984) from p. 503 of Ref.~\cite{Gradshteyn})
\begin{eqnarray}
\label{p1wB}
W^{(00)}_{\lambda}(s,\,q) &=& 2\frac{\mathcal{A}^{\2}(\lambda)}{\Gamma(\lambda)}\mathcal{D}^{\lambda -1}_{(s)}\,f(s,\,q)\\
\label{p2wB}
W^{(11)}_{\lambda}(s,\,q) &=& - 4(\lambda - 1)\mathcal{A}^{\2}(\lambda)\,\left[\frac{2}{\Gamma(\lambda-1)}+\frac{\mbox{coth}(2s)}{\Gamma(\lambda)}\frac{d\,}{ds}\right]\mathcal{D}^{\lambda -2}_{(s)}\,f(s,\,q),
\end{eqnarray}
with the differential operator
\begin{equation}
\mathcal{D}^{\lambda}_{(s)} \equiv \left(\frac{(-1)}{\sinh(2s)}\frac{d\,}{ds}\right)^{\lambda},
\end{equation}
and
\begin{equation}
f(s,\,q) 
=\frac{2}{\pi} \int^{\infty}_{0}dy\,\frac{\cos(2\,q \,y)}{\cosh(2s)+\cosh(2y)}
= \frac{\sin(2\,q \,s)}{\sinh(2s)\,\sinh(\pi q)}.
\end{equation}

\begin{figure}
\includegraphics[scale=0.48]{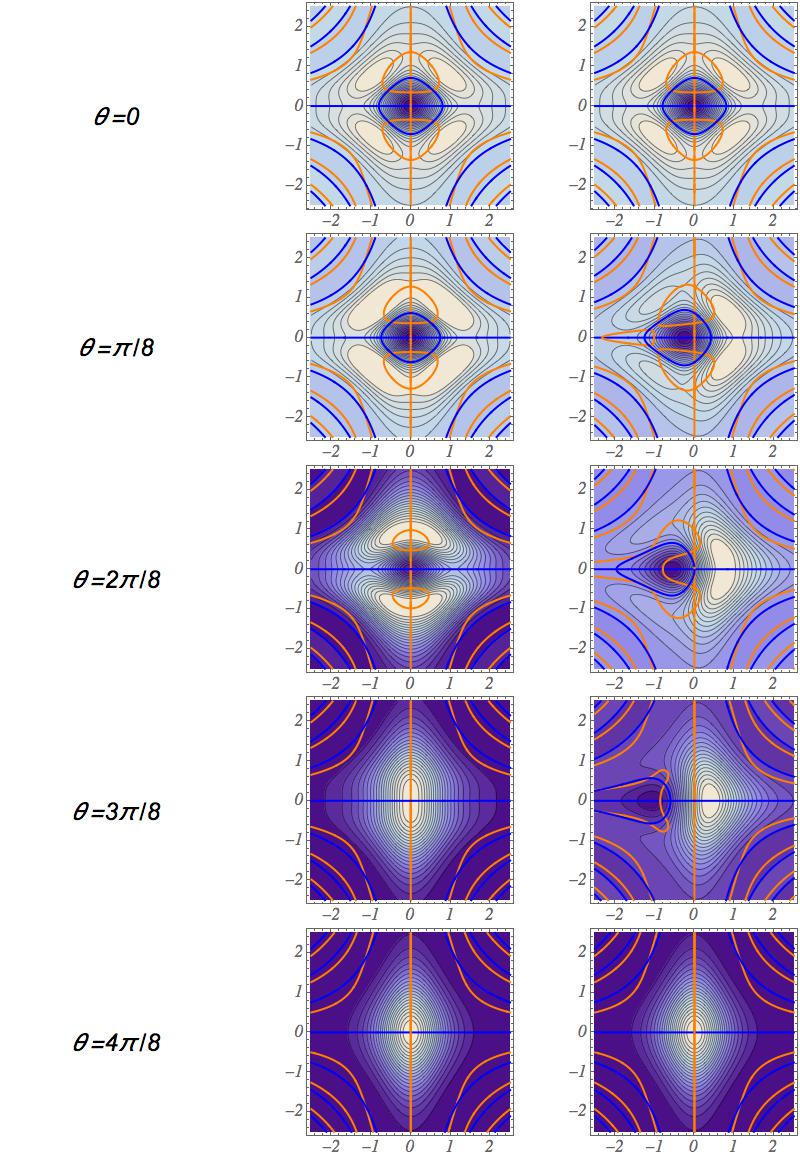}
\renewcommand{\baselinestretch}{.85}
\caption{
(Color online) Wigner function contour plot profile in the $s-q$ plane, $W(s,\,q)$, for mixed states (left column), $\sin^{\2}(\theta)\, W^{00}_{2} + \cos^{\2}(\theta)\, W^{11}_{2}$, and quantum superpositions (right column), $W^{C}_{2} = \sin^{\2}(\theta)\, W^{00}_2 + \cos^{\2}(\theta)\, W^{11}_2 +(1/2)\sin(2\theta) W^{(1\leftrightarrow0)}_2(\varphi \rightarrow 0)$, between ground and first excited states for the PT potential with $\lambda = 2$, for mixing angles $\theta = n \pi/8$ with $n = 0,\,1,\, \dots,\,4$, from top to bottom.
Blue contour lines are for $J_s = 0$ and orange contour lines are for $J_q = 0$.
Blue (orange) contour lines define the bounds for the reversion of the Wigner flow along the $s(q)$ direction.
The observed blue-orange intersections define vortices and saddle-points which introduce the quantum fluctuations.}
\label{FiguraAAA}
\end{figure}

Considering the quantum superposition described by $\psi^C_{\lambda}(s) = \sin(\theta)\, \psi^{0}_{\lambda}\,\exp(-i\varphi) + \cos(\theta)\,\psi^{1}_{\lambda}\exp(+i\varphi)$ -- when replaced by the phase-space representation -- it results in the complete Wigner function given by $$W^{C}_{\lambda} = \sin^2(\theta)\,W^{(00)}_{\lambda} + \cos^2(\theta)\,W^{(11)}_{\lambda} + \frac{\sin(2\theta)}{2} W^{(1\leftrightarrow0)}_{\lambda},$$
which involves a mixed Wigner function contribution,
\begin{eqnarray}
\label{p1wBBB}
W^{(1\leftrightarrow0)}_{\lambda}(s,\,q) &=& \exp(-2i\varphi)W^{(10)}_{\lambda}(s,\,q)+\exp(+2i\varphi)W^{(01)}_{\lambda}(s,\,q)\nonumber\\
 &=& 2^{\frac{5}{2}}(\lambda - 1)^{\frac{1}{2}}\frac{\mathcal{A}^{\2}(\lambda)}{\Gamma(\lambda)}
\left[\cos(\varphi)\,\sinh(s)\,\mathcal{D}^{\lambda -1}_{(s)}\,g(s,\,q) \right.\nonumber\\
&&\qquad\qquad\qquad\qquad\qquad\qquad\qquad\qquad\left.+ \sin(\varphi)\,\cosh(s)\,\mathcal{D}^{\lambda -1}_{(s)}\,h(s,\,q)\right],\qquad
\end{eqnarray}
with (cf. Ref.~\cite{Gradshteyn})
\begin{equation}
g(s,\,q) 
=\frac{2}{\pi} \int^{\infty}_{0}dy\,\frac{\cosh(y)\,\cos(2\,q \,y)}{\cosh(2s)+\cosh(2y)}
= \frac{\cos(2\,q \,s)}{2\cosh(2s)\,\cosh(\pi q)},
\end{equation}
and
\begin{equation}
h(s,\,q) 
=\frac{2}{\pi} \int^{\infty}_{0}dy\,\frac{\sinh(y)\,\sin(2\,q \,y)}{\cosh(2s)+\cosh(2y)}
= \frac{\sin(2\,q \,s)}{\sinh(2s)\,\cosh(\pi q)}.
\end{equation}
The phase-space representation of the above results for a mixed state and a pure state quantum superposition involving ground and first excited states is depicted in Fig.~\ref{FiguraAAA}: 
the left panels show the Wigner profile for the rank 2 mixed state described by $\sin^{\2}(\theta)\, W^{00}_{2} + \cos^{\2}(\theta)\, W^{11}_{2}$; the right panels show the Wigner function for the quantum superposition $W^{C}_{2} = \sin^{\2}(\theta)\, W^{00}_2 + \cos^{\2}(\theta)\, W^{11}_2 +(1/2)\sin(2\theta) W^{(1\leftrightarrow0)}_2(\varphi \rightarrow 0)$, in both cases, for a PT potential with $\lambda = 2$.
The mixing angle $\theta$ is set equal to $n \pi/8$ with $n = 0,\,1,\, \dots,\,4$, from top to bottom panels and, for the pure state quantum superposition (right panels), $\theta$ equivalently works as a time evolving variable which performs a continuous quantum oscillation scheme of a PT two-level system.
At this point, besides the quasi-Gaussian profile exhibited by the ground-state $W^{00}_2$, its pertinent to notice that the choice of increasing values for $\lambda$ would not affect qualitatively neither the Wigner function profiles nor their related dynamics.

More interesting aspects of the dynamics of quantum superpositions \cite{Steuernagel3,Ferraro11} can be revealed by the behavior the Wigner function when it is cast in the form of a flow field $\vec{J}(s,\,q;\,\tau) = J_s\,\hat{n}_s + J_q\,\hat{n}_q$, where $\tau$ is the previously introduced dimensionless time variable, $\tau = 2\varepsilon t/\hbar$, and $\hat{n}_\kappa$, with $\kappa = s,\,q$, are unitary vectors, $\vert\hat{n}_{\kappa}\vert = 1$, for the phase-space coordinates, $s$ and $q$. It describes the vectorial flux of the quasi-probability density in the phase-space \cite{Donoso12,Domcke}, where $\tau$ is identified with the above introduced phase parameter, $\varphi = (\lambda - 1/2) \tau$.
The equivalent quantum Liouville equation is written in terms of a dimensionless version of the continuity equation in the phase-space \cite{Case,Ballentine,Steuernagel3} as
\begin{equation}
\frac{\partial W^C_{\lambda}}{\partial \tau} + \frac{\partial J_s}{\partial s}+\frac{\partial J_q}{\partial q} =
\frac{\partial W^C_{\lambda}}{\partial \tau} + \vec{\nabla}\cdot \vec{J} =0,
\label{eqn51}
\end{equation}
where
\begin{eqnarray}
J_s(s,\,q;\,\tau)&=& q\,W^{C}_{\lambda}(s,\,q;\,\tau),\nonumber\\
J_q(s,\,q;\,\tau)&=& \frac{1}{2}\lambda(\lambda+1)\sum_{k=0}^{\infty} \frac{(-1)^k}{2^{2k}}\frac{1}{(2k+1)!} \,\left( \left(\frac{\partial~}{\partial s}\right)^{2k+1}\hspace{-.5cm}\mbox{sech}^{\2}(s)\right)\,\left(\frac{\partial~}{\partial q}\right)^{2k}\hspace{-.3cm}W^C_{\lambda}(s,\,q;\,\tau).\quad
\label{eqn500}
\end{eqnarray}
The above quantities, once related to the Wigner function for $\lambda = 2$, $W^C_{2}$, are also depicted in Fig.~\ref{FiguraAAA}, in particular, for $J_s = 0$ (blue contour lines), and $J_q = 0$ (orange contour lines).

\begin{figure}
\vspace{-1 cm}
\includegraphics[scale=0.4]{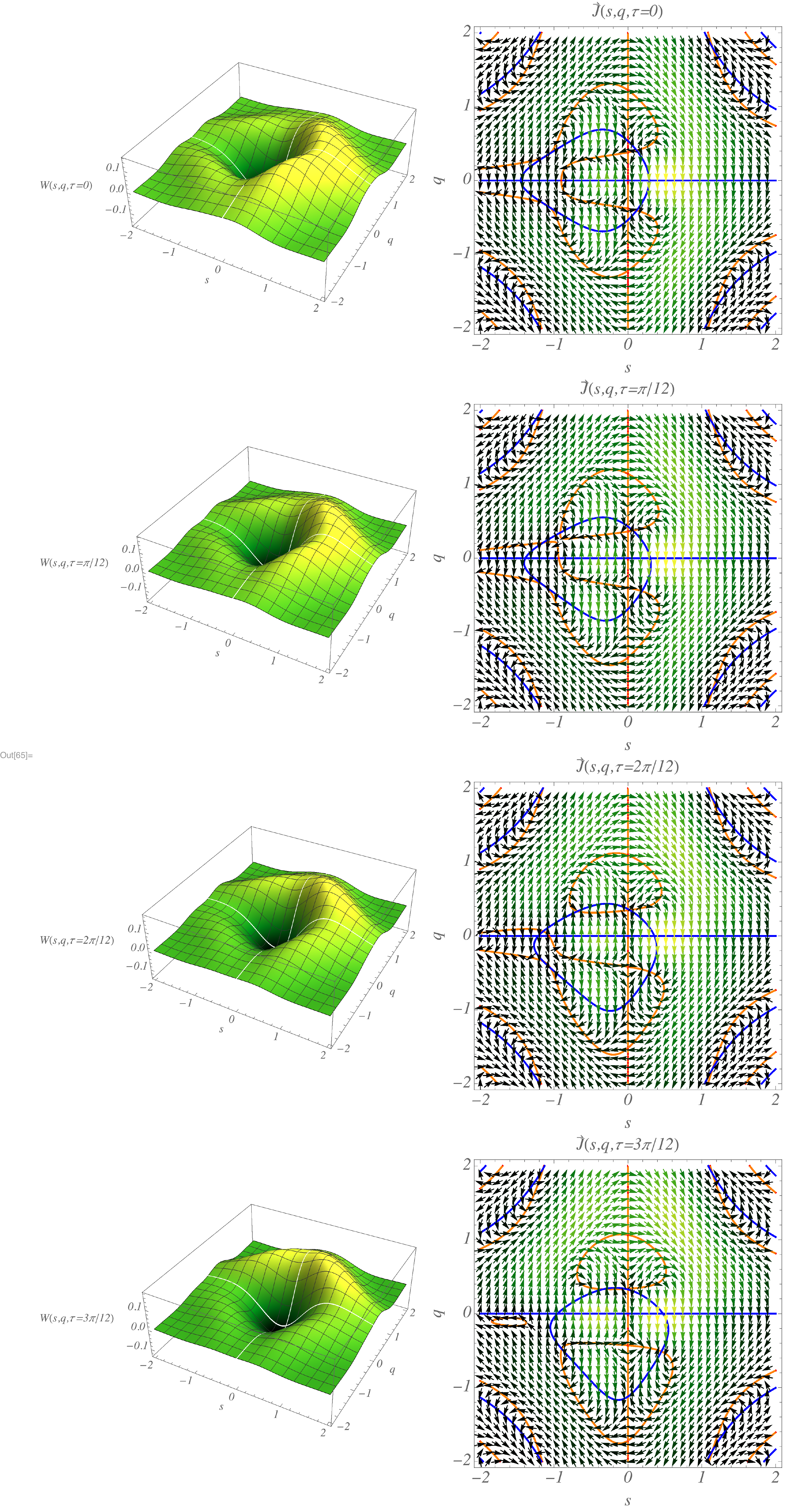}
\renewcommand{\baselinestretch}{.85}
\caption{
(Color online) 
Time evolution of the Wigner function (left) and corresponding Wigner flow, $\vec{J}/\vert\vec{J}\vert$,
for the ground and first excited state quantum superposition, $W^{C}_{2}$, with $\theta = \pi/6$, and $\tau =n \pi/12$ with $n = 0,\,1,\,2,$ and $3$, from top to bottom. 
Field and line notations are in correspondence with Fig.~\ref{FiguraAAA}.
}\label{FiguraCCC}
\end{figure}

\begin{figure}
\includegraphics[scale=0.54]{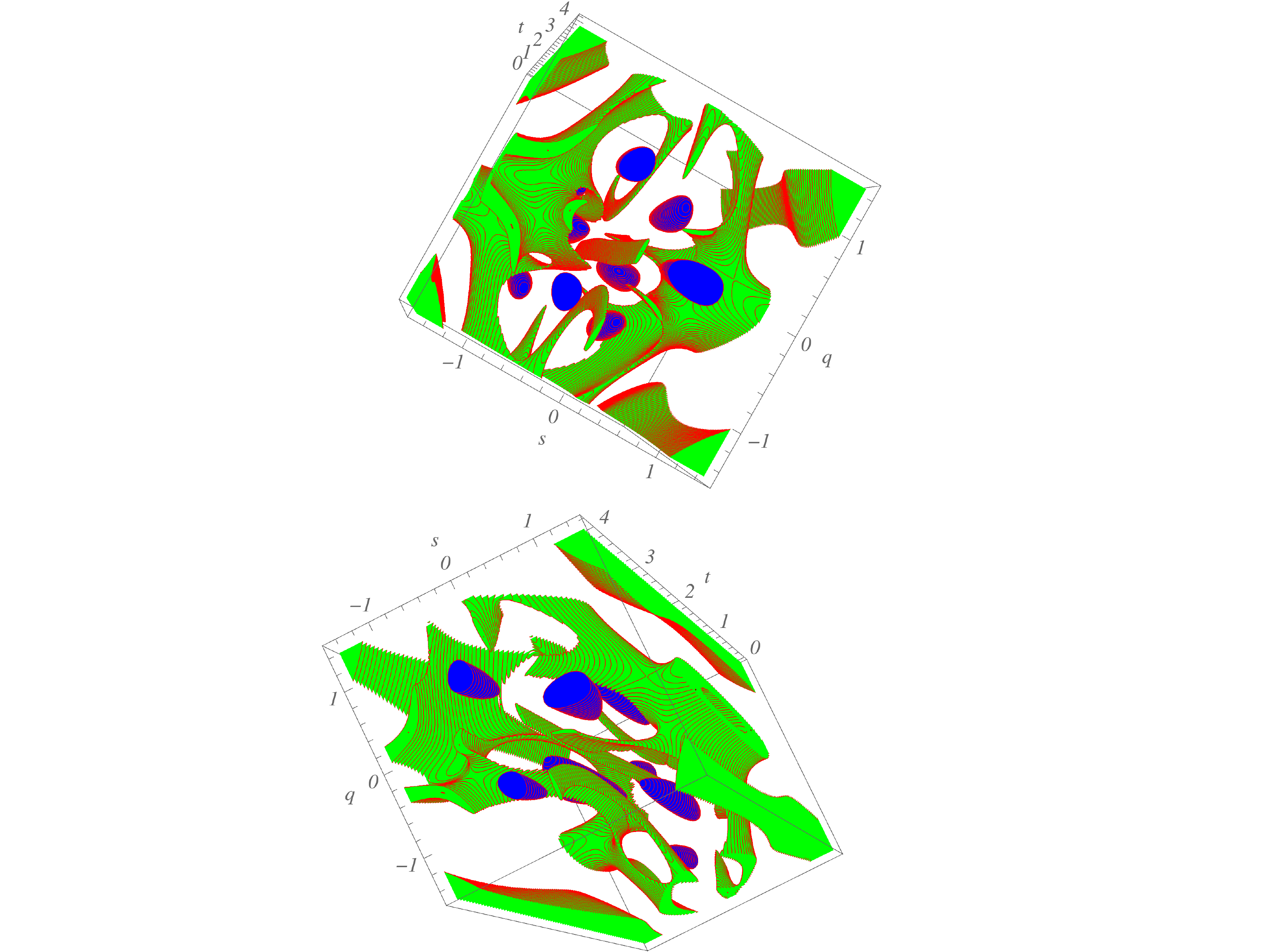}\hspace{-4cm}
\renewcommand{\baselinestretch}{.85}
\caption{
(Color online) 
Slice plot scheme for the time evolution of domain walls (blue bubbles) and quantum critical points (green islands) for the Wigner function in correspondence with Fig.~\ref{FiguraCCC}. The pictures show an increased time interval in order to illustrate the continuous evolution of critical points and domain walls.}
\label{FiguraDDD}
\end{figure}

In fact, a straightforward picture of intrinsic quantum effects can be patterned from the contributions due to $k\leq 1$ for $J_q(s,\,q;\,\tau)$ at Eq.~(\ref{eqn500}).
Such quantum fluctuations are more clearly depicted in Figs.~\ref{FiguraCCC} and \ref{FiguraDDD}.
Fig.~\ref{FiguraCCC} shows the normalized Wigner flow (right panels) for the two-level oscillating system, $W^{C}_{2}$, in correspondence with its surface plot (merely illustrative) representation (left panels).
The pictures are for $\lambda =2$, $\theta = \pi/6$, and $\tau =n \pi/12$ with $n = 0,\,1,\,2,$ and $3$, from top to bottom.
As it can be noticed, blue contour lines, for $J_s = 0$, and orange contour lines, for $J_q = 0$, define the bounds for the reversion of the Wigner flow along the corresponding $s$ and $q$ directions. Typical quantum effects are identified by the intersection of blue and orange contour lines which simultaneously sets $J_{s}=J_{q}=0$ and defines vortices and saddle-points which introduce the quantum local effects that perturb the classical pattern.
The time evolving scheme from Fig.~\ref{FiguraCCC} allows one to identify the smooth evolution of the domain wall regions and of the creation/annihilation of quantum distortions (i.e. $J_q = J_s = 0$ for intersecting blue-orange lines). Correspondently, the domain wall regions associated to the dominating components of the Wigner flux, $\vec{J}/\vert\vec{J}\vert$, evolve continuously in time, as it is depicted by blue bubble regions in the slice plot scheme from Fig.~\ref{FiguraDDD}, which follows a parallel illustrative proposal.
In this case, the continuous time evolution for vortices and saddle-points is identified by the green islands of the slice plots. As they are noticeable in Fig.~\ref{FiguraCCC}, the local quantum effects compensate each other when an integrated global view of the Wigner flux is considered, i.e. when two vortices of opposite circulating (or divergence) number match each other along the time evolution. They mutually self-annihilate and approach the background classical pattern.

\subsection{Non-Liouvillian effects for time-oscillating two-level systems}

In particular, the quantum perturbations over the classical pattern can be discussed in terms of a quantifier for the non-Liouvillian effects exhibited by the quantum system.
For the classical pattern, the phase-space coordinate vector $\vec{\xi} = (s,\,q)$ has the corresponding classical Hamiltonian velocity identified by 
$\dot{\vec{\xi}} = \vec{v}_{\xi} = (v_s,\,v_q)$. The flow field is thus identified by $\vec{J} = \vec{v}_{\xi}\,W$, with $v_s = \dot{s} = q$ and $v_q = \dot{q} = -\partial V/\partial s$, that satisfy the divergenceless Liouvillian condition, $\vec{\nabla}_{\xi} \cdot \vec{v}_{\xi} = 0$.
On the other hand, for the quantum analysis, when one assumes the complete current expansion from Eqs.~(\ref{eqn500}), one can identify $\vec{J}$ with $\vec{u}\,W$, and a typical non-Liouvillian \cite{Liouvillian} flow can be described by $\vec{\nabla}_{\xi} \cdot \vec{u} \neq 0$ since one has
\begin{equation}
\vec{\nabla}_{\xi} \cdot \vec{u} = \frac{W\, \vec{\nabla}\cdot \vec{J} - \vec{J}\cdot\vec{\nabla}W}{W^2},
\label{eqn59}
\end{equation}
where $\vec{\nabla}\cdot\vec{J} = W\,\vec{\nabla}\cdot\vec{u}+ \vec{u}\cdot \vec{\nabla}W$.
The condition that sets $\vec{\nabla}_{\xi} \cdot \vec{u} \neq 0$ is very helpful in identifying (approximated) Liouvillian-like trajectories in the phase-space and it has been used to quantify the correspondence between quantum and classical descriptions \cite{Our01,Our02}.

\begin{figure}
\includegraphics[scale=0.45]{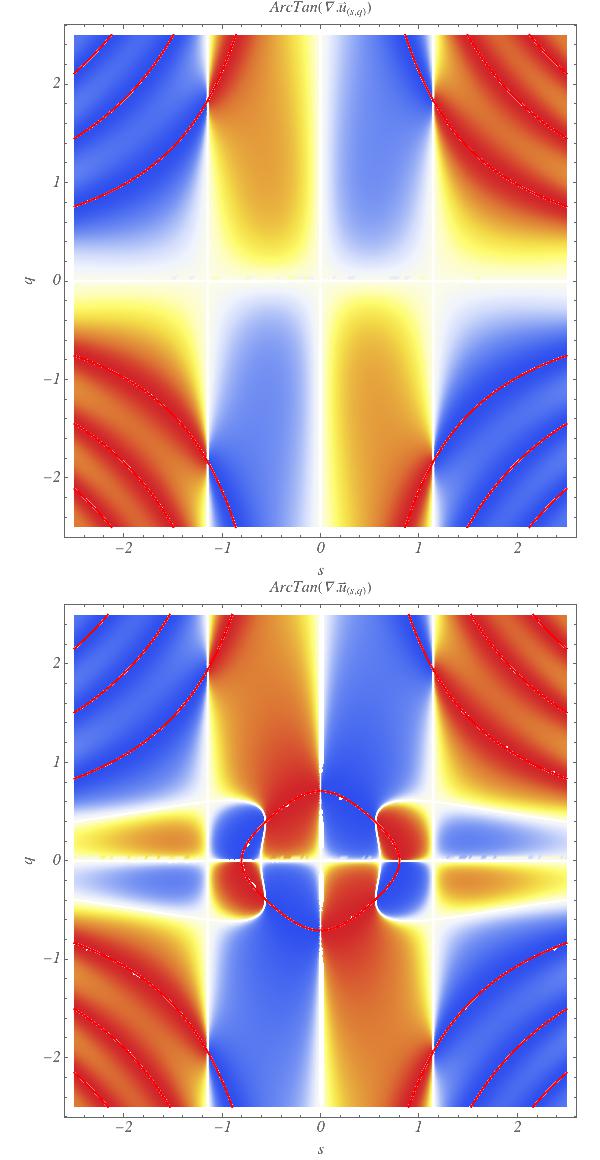}
\renewcommand{\baselinestretch}{.85}
\caption{
(Color online)
Liouvillian quantifier -- for ground (upper panel) and first excited (lower panel) states -- parameterized by $\mbox{arctan}(\vec{\nabla} \cdot \vec{w})$ in the phase-space ($s - q$ plane). White-lines are for $\vec{\nabla} \cdot \vec{u} = 0$ red-lines are for $W^C_2=0$. The {\em TemperatureMap} color scheme -- from blue-regions where $\mbox{arctan}(\vec{\nabla} \cdot \vec{u})\sim -1$, to red-regions where $\mbox{arctan}(\vec{\nabla} \cdot \vec{u})\sim +1$ -- reinforces the approximated Liouvillian behavior for white-regions (more evinced for the ground-state).}
\label{FiguraBBB}
\end{figure}

Fig.~\ref{FiguraBBB} shows the results for
the corresponding Liouvillian quantifier parameterized by $\mbox{arctan}(\vec{\nabla} \cdot \vec{u})$ in the phase-space ($s - q$ plane).
The results can be interpreted according to a color scheme -- from blue-regions where $\mbox{arctan}(\vec{\nabla} \cdot \vec{u})\sim -1$, to red-regions where $\mbox{arctan}(\vec{\nabla} \cdot \vec{u})\sim +1$ -- which reinforces the approximated Liouvillian behavior for white-regions (more evinced for the ground-state), where quantum effects are typically suppressed.
In fact, the plots only provide a qualitative view of the non-Liouvillian pattern, which increases for excited states, with respect to the ground one.
The above results resume a qualitative approach for identifying quantum effects into PT quantum state configurations and can be extended to more involved configurations.
In the following, the quantitative aspects of the quantum information profile associated to the above analytical description will be developed and compared with general descriptions of quantum to classical correspondence frameworks.

\section{Two-level system information profile} 

Characterizing quantum states with the positivity of Wigner functions \cite{Alba16,Genoni00} and distinguishing their Gaussian and non-Gaussian (information) profiles \cite{Genoni01,Genoni02} are often considered in several classification protocols according to classical/non-classical paradigms.
Preliminary results also indicate that quantum non-Gaussianity can only be produced by means of highly non-linear processes \cite{Genoni00}, which involve, for instance, the anharmonic dynamics of PT potentials.
Considering the analytical results from Sec.~III, through the computation of some different measures of non-Gaussianity for hyperbolic PT quantum superpositions involving ground and first-excited states, one is able to identify and quantify some associated two-level system (non-)Gaussian information profiles, as it shall be demonstrated in the following sections.

\subsection{Wigner negativity, kurtosis and entropic non-Gaussianity}
			
Negativities of the Wigner function associated with a given quantum state can be classified as a pattern of non-classicality \cite{Alba16}. For the hyperbolic PT ground and first-excited pure states described by Eqs.~\eqref{p1wB} and \eqref{p2wB} -- and, of course, for their associated mixed states -- the negativity of the Wigner function is null. 
Although some non-vanishing values of negativity grow up for quantum superpositions, this fact does not introduce any relevant information about the Gaussianity of the corresponding pure states.

\begin{figure}[h]
\includegraphics[scale=1.05]{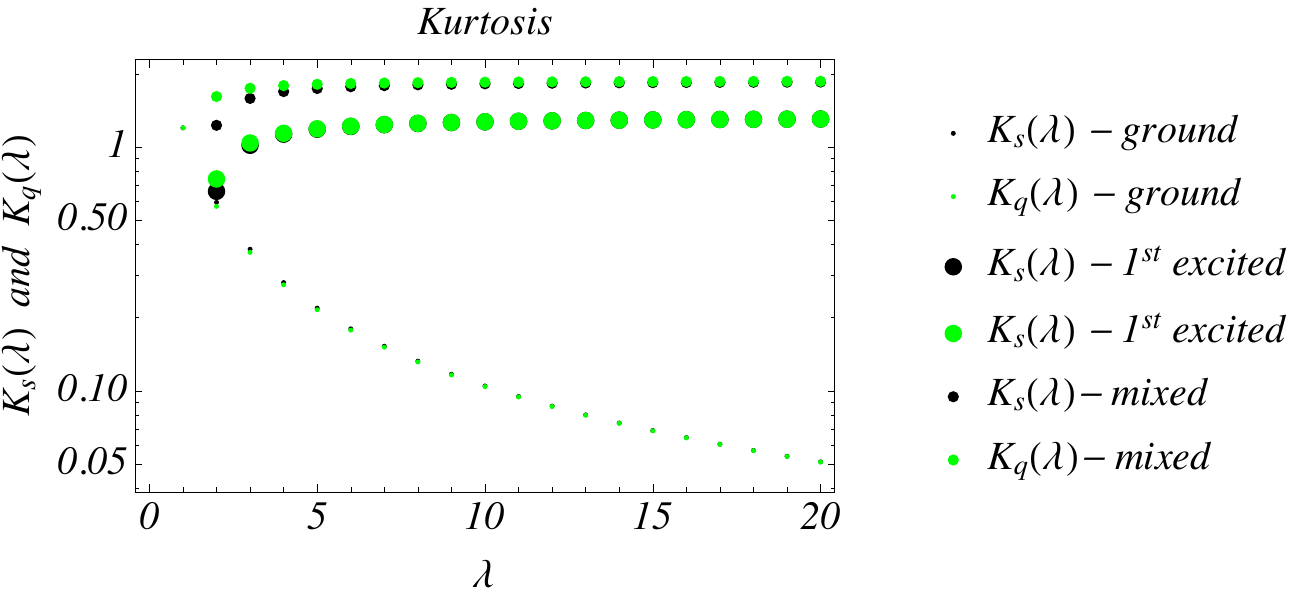}
\renewcommand{\baselinestretch}{.85}
\caption{
(Color online) Kurtosis of position (dark black circles) and momentum (light green circles) for ground (small size) and first excited (large size), as well as for maximized ($\theta = \pi/4$) mixed states (medium size) as function of the $\lambda$ quantum number.}
\label{Kurtosis}
\end{figure}

Otherwise, the first ever considered measure of non-Gaussianity, the (excess) kurtosis, given by the regularized fourth-order cumulant
\begin{equation}
\mathcal{K}_{\xi} = \frac{\langle \xi ^4 \rangle - \langle \xi \rangle^4}{(\langle \xi ^2 \rangle - \langle \xi \rangle^2)^2} - 3 \qquad \mbox{with} \quad \xi = s,\, q,
\end{equation}
can be straightforwardly calculated as to give the results depicted in Fig.~\ref{Kurtosis} for coordinate and momentum variables, $s$ and $q$.
It is interesting to notice that, according to the (excess) kurtosis criterium, for deeper quantum wells ($\lambda \gg 1$), the ground-state can be approached by Gaussian ones since the PT quantum well approaches the harmonic oscillator one.
For first excited states and for maximized statistical mixtures, where the mixing angle is given by $\theta = \pi/4$, the increasing values of $\lambda$ stabilizes the kurtosis respectively around $\mathcal{K} = 1.68$ and $1.13$, for which Gaussian approaches, in the context of information measures, are still implementable.
Due to the parity symmetry of the Wigner functions, results for mixed states and quantum superpositions are equivalent in this case.

\begin{figure}[h]
\includegraphics[scale=0.55]{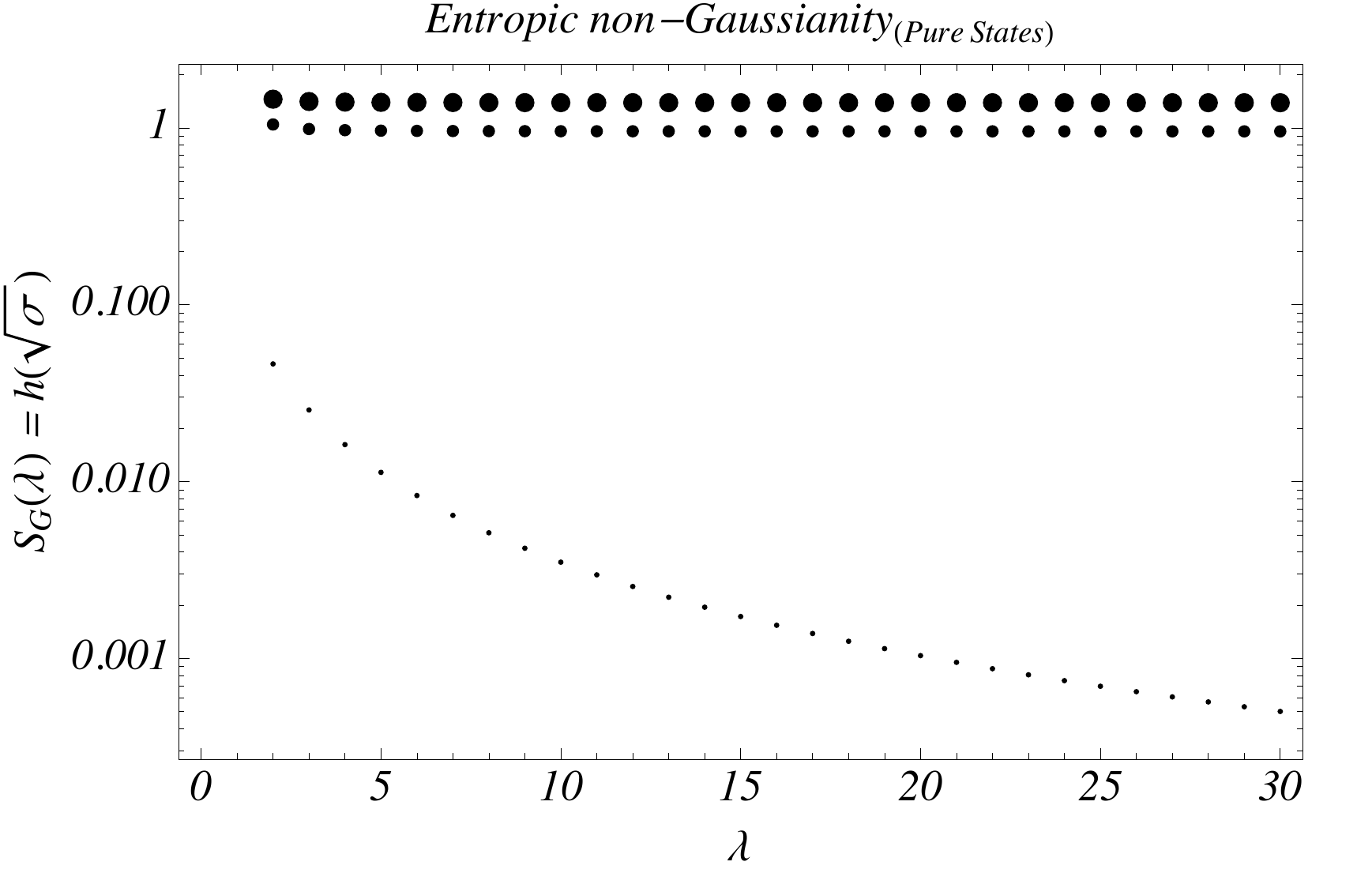}
\renewcommand{\baselinestretch}{.85}
\caption{(Color online) Entropic non-Gaussianity for ground (small size) and first excited (large size) states, as well as for a pure state quantum superposition with $\theta = \pi/4$ (medium size) as function of the $\lambda$ quantum number.}
\label{Negen}
\end{figure}

Such a behavior is ratified by a more involved quantifier, the entropic non-Gaussianity, which is given in terms of the difference between the exact Shannon entropy for the Wigner function, $S(\omega(\xi))$, and a Gaussian approach for the von Neumann entropy \cite{Olivares,Holevo,Serafini},
\begin{equation}
S_G(\omega(\xi)) = h(\sqrt{\sigma})
\label{nene}
\end{equation}
with $h(z) = (z+\frac{1}{2}) \ln(z+\frac{1}{2}) - (z-\frac{1}{2}) \ln(z-\frac{1}{2})$, and where $\sigma$ is the determinant of the $2 \times 2$ covariance matrix with elements identified by
\begin{equation}\label{covar}
\sigma_{ij}(\omega) = \frac{1}{2} \langle \{ \xi_i,\,\xi_j\}\rangle_{\omega} - \langle \xi_i\rangle_{\omega}\,\langle \xi_j\rangle_{\omega}
\qquad \mbox{with} \quad \xi_{1,2} = s,\, q.
\end{equation}
For pure states, with $S(\omega(\xi))=0$, the entropic non-Gaussianity is straightforwardly given by $S_G(\omega(\xi))$ from Eq.~\eqref{nene}, as shown in Fig.~\ref{Negen}.
Despite the quantitative similarities with the results from Fig.~\ref{Kurtosis}, there is an inverted correspondence between first excited states and quantum superposition results which, in this case, stabilizes respectively around $\mathcal{K} = 0.95$ and $\mathcal{K} = 1.37$.

The above results work coincidently fine in describing the non-Gaussian behavior of the ground-state at lower values of $\lambda$ and its corresponding ``approximated'' harmonic oscillator Gaussian behavior for increasing values of $\lambda$.
In the latter case, the PT potential approaches an harmonic oscillator potential profile.
At this point, it is convenient fo assert that defining quantitative measures of non-classicality associated to non-Gaussianity \cite{Olivares} is still an open issue.
For this reason, a more elaborated explanation for the opposite saturated non-Gaussian pattern for first excited (large size circles) and mixed (medium size circles) states in Figs. \ref{Kurtosis} and \ref{Negen} cannot be provided by this short analysis.
Even if non-Gaussianity cannot be considered as a quantum resource of practical relevance \cite{Alba18} and
a faithful measurement criterium has not been established yet, non-Gaussianity can be generated by classical randomness readily available from an operational point of view \cite{Park18}.
Therefore, the large number of theoretical tools developed to assess quantum non-Gaussian states, rule out Gaussian mixtures and detect genuinely quantum non-Gaussian states, clearly deserves more careful investigations.
Likewise, when the discussion of the quantum information profile is extended to a two-level system framework, Gaussian and non-Gaussian more global criteria must be considered.

\subsection{Quantum separability of two-level systems}

For a bipartite state written as
\begin{equation} 
\label{state}
\vert \Psi \rangle = \frac{1}{\sqrt{2}} \left( \vert 0_{_A} 1_{_B} \rangle + \vert 1_{_A} 0_{_B} \rangle \right),
\end{equation}
where $\vert 0\rangle$ and $\vert 1\rangle$ denote PT ground and first excited states of $A$ and $B$ subsystems, the continuous variable entanglement in this bipartite system is more conveniently introduced through the density matrix written as
\begin{equation}
\vert\Psi\rangle\langle\Psi\vert = \frac{1}{2}\left( \vert 0_{_A} 1_{_B} \rangle\langle 0_{_A} 1_{_B}\vert + \vert 0_{_A} 1_{_B} \rangle\langle 1_{_A} 0_{_B}\vert + \vert 1_{_A} 0_{_B} \rangle\langle 0_{_A} 1_{_B}\vert +\vert 1_{_A} 0_{_B} \rangle\langle 1_{_A} 0_{_B}\vert \right),
\end{equation}
which is identified by the $2+2$-dim phase space variable Wigner function
\begin{equation}\small \label{wigner-def}
W(s_{_A}, s_{_B};\,q_{_A}, q_{_B}) = \frac{1}{\pi^2}\int_{-\infty}^{+\infty}\hspace{-.4cm}dy_{_A}\,\int_{-\infty}^{+\infty}\hspace{-.4cm}dy_{_B} \langle s_{_A} + y_{_A}, s_{_B} + y_{_B} \vert\Psi\rangle\langle\Psi \vert s_{_A}-y_{_A},\, s_{_B}-y_{_B} \rangle e^{2i(p_{_A} y_{_A}+p_{_B} y_{_B})}.
\end{equation}\normalsize

Using the previous results, and denoting the quantum number $\lambda$, it is straightforward to show that the above identified Wigner function takes the form of
\begin{eqnarray} 
\label{wigner}
W_{\lambda} (s_{_A}, s_{_B};\,q_{_A}, q_{_B}) &=& \frac{1}{2}\bigg{[}W^{(00)}_{\lambda}(s_{_A},q_{_A})\, W^{(11)}_{\lambda}(s_{_B},q_{_B})+ 
W^{(11)}_{\lambda}(s_{_A},q_{_A}) \,W^{(00)}_{\lambda}(s_{_B},q_{_B}) \nonumber\\&&\qquad\qquad\qquad\qquad\qquad+ \left(W^{(01)}_{\lambda}(s_{_A},q_{_A})\, W^{(10)}_{\lambda}(s_{_B},q_{_B}) +\, h.c.\right)\bigg{]},\quad
\end{eqnarray}
where $W^{(00)}_{\lambda}(s,\,q)$ and $W^{(11)}_{\lambda}(s,\,q)$ are respectively given by Eqs.~\eqref{p1wB} and \eqref{p2wB}, and 
\begin{eqnarray} 
\label{wigner2}
W^{(10)}_{\lambda}(s,\,q) &=& W^{*(01)}_{\lambda}(s,\,q) \nonumber\\
&=& \left(8(\lambda-1)\right)^{\frac{1}{2}}\frac{\mathcal{A}^{\2}(\lambda)}{\Gamma(\lambda)}
\left[\sinh(s)\,\mathcal{D}^{\lambda -1}_{(s)}\,g(s,\,q) + i\,\cosh(s)\,\mathcal{D}^{\lambda -1}_{(s)}\,h(s,\,q)\right],
\end{eqnarray}
from which, even considering the mentioned non-Gaussian profile, one will be able to obtain analytical expressions for second order moments in position and momentum, namely a helpful task in building the continuous variable entanglement tests \cite{Bertolami2017}.
By noticing that $W^{(00)}_{\lambda}(s,\,q)$ and $W^{(11)}_{\lambda}(s,\,q)$ are typically even parity functions in $s$ and $q$, and that $W^{(10)}_{\lambda}(s,\,q)$ has a real (imaginary) contribution with even parity in $q$ ($s$) and odd parity in $s$ ($q$), the computation of first and second order moments can be further simplified.
By identifying each contribution of averaged integration of a generic operator $\mathcal{O}(s,\,q)$ with, 
\begin{equation}
\langle \mathcal{O}(s,\,q)\rangle^{(ij)} = \int_{-\infty}^{+\infty}\hspace{-.4cm}ds\, \int_{-\infty}^{+\infty}\hspace{-.4cm}dq\,\,\mathcal{O}(s,\,q)\,W^{(ij)}_{\lambda}(s,\,q),
\label{integral}
\end{equation}
where $i,j = 0,\,1$, one obtains that all the first moments are null (due to previous parity considerations), and also that $\langle q^{\2}\rangle^{10} = \langle s^{\2}\rangle^{10} = \langle s q\rangle^{11}= \langle s q\rangle^{00} = 0$, for separated subsystems $A$ or $B$.
From the analytical evaluation of integrals like (\ref{integral}), the $A$ and $B$ separated non-vanishing contributions to second moment values result into 
\begin{eqnarray} 
\label{averaged2}
\langle s^2_{_{A,B}}\rangle^{(00)} &=&\frac{1}{2}\frac{d^{\2}}{d\lambda^{\2}}\ln\left(\Gamma(\lambda)\right),\\
\langle s^2_{_{A,B}}\rangle^{(11)} &=&\frac{1}{2}\left[\frac{d^{\2}}{d\lambda^{\2}}\ln\left(\Gamma(\lambda)\right)+\frac{2\lambda -1}{(\lambda -1)^{\2}}\right],\\
\langle q^2_{_{A,B}}\rangle^{(00)} &=&\frac{\lambda^{\2}}{2\lambda+1},\\
\langle q^2_{_{A,B}}\rangle^{(11)} &=&\frac{(\lambda-1)(3\lambda+1)}{2\lambda+1}.
\end{eqnarray}
In addition, some relevant contributions due to mixed subsystem components from $A$ and $B$ can be computed as to result into non-vanishing values for $\langle s_{_A} s_{_B}\rangle$, $\langle q_{_A} q_{_B}\rangle$, $\langle s_{_A} q_{_B}\rangle$ and $\langle q_{_A} s_{_B}\rangle$.
Such averaged products can be decomposed into separated contributions from $A$ and $B$ given in terms of the products between non-vanishing one-mode averaged values of $s$ and $q$, $\langle s\rangle_{A(B)}$ and $\langle q\rangle_{B(A)}$ which, due to the parity properties of $W^{(ij)}$, only involve contributions from $W^{(10)}_{\lambda}$ as to give 
\begin{eqnarray} 
\langle s_{_{A,B}}\rangle^{(10)} &=&\left(\frac{\lambda-1}{2}\right)^{\frac{1}{2}}\frac{\Gamma^{\2}\left(\lambda-\frac{1}{2}\right)}{\Gamma^{\2}(\lambda)},\\
\langle q_{_{A,B}}\rangle^{(10)} &=&i\,\left(\lambda-\frac{1}{2}\right)\left(\frac{\lambda-1}{2}\right)^{\frac{1}{2}}\frac{\Gamma^{\2}\left(\lambda-\frac{1}{2}\right)}{\Gamma^{\2}(\lambda)},
\label{averaged}
\end{eqnarray}
from which, through the Hermitian properties of the Wigner functions (cf. Eq.~\eqref{wigner}), some simple manipulations suppress the contribution due to the product $\langle s_{A(B)} q_{B(A)}\rangle$.

Extending the notation for the covariance matrix definition from \eqref{covar} to a two-mode (bipartite) quantum system, with the phase-space variables identified by $\xi_{1,2,3,4} = s_{_A},\, q_{_A},\,s_{_B},\,q_{_B}$, all the above results are used to build the covariance matrix of the bipartite state from (\ref{wigner-def}),
\begin{equation} \label{covariance-state}
{\sigma} =
\left(
\begin{array}{cccc}
\langle s^2_{_A}\rangle & \frac{1}{2} \langle\{s,q\}_{_A}\rangle & \langle s_{_A}\,s_{_B}\rangle & \langle s_{_A}\,q_{_B}\rangle \\
 \frac{1}{2} \langle\{s,q\}_{_A}\rangle & \langle q^2_{_A}\rangle &\langle q_{_A}\,s_{_B}\rangle & \langle q_{_A}\,q_{_B}\rangle \\
\langle s_{_A}\,s_{_B}\rangle & \langle s_{_A}\,q_{_B} \rangle& \langle s^2_{_B}\rangle &\frac{1}{2}\langle \{s,q\}_{_B}\rangle \\
\langle q_{_A}\,s_{_B}\rangle & \langle q_{_A}\,q_{_B} \rangle& \frac{1}{2} \langle\{s,q\}_{_B}\rangle&\langle q^2_{_B}\rangle \\
\end{array}
\right)
=
\left(
\begin{array}{cccc}
\alpha_s & 0 & \gamma_s & 0 \\
0 &\alpha_q& 0 & \gamma_q\\
\gamma_s & 0 & \alpha_s&0\\
0& \gamma_q& 0&\alpha_q
\end{array}
\right),
 \end{equation}
where $\alpha_\kappa = \left(\langle \kappa^2_{_{A,B}}\rangle^{(00)} +\langle \kappa^2_{_{A,B}}\rangle^{(11)}\right)/2$ and $\gamma_\kappa =\left(\langle \kappa_{_A}\rangle^{(10)}\langle \kappa_{_B}\rangle^{(01)} +\langle \kappa_{_A}\rangle^{(01)}\langle \kappa_{_B}\rangle^{(10)}\right)/2$, with $\kappa = s,\, q$, so as to result in
\begin{eqnarray} 
\alpha_s &=& \frac{1}{2}\left[\frac{d^{\2}}{d\lambda^{\2}}\ln\left(\Gamma(\lambda)\right)+\frac{2\lambda -1}{2(\lambda -1)^{\2}}\right],\\
\alpha_q &=& \frac{4\lambda^{\2}- 2 \lambda - 1}{4\lambda+2},\\
\gamma_s &=& \left(\frac{\lambda-1}{2}\right)\frac{\Gamma^{\4}\left(\lambda-\frac{1}{2}\right)}{\Gamma^{\4}(\lambda)},\\
\gamma_q &=& \left(\lambda-\frac{1}{2}\right)^{\2}\left(\frac{\lambda-1}{2}\right)\frac{\Gamma^{\4}\left(\lambda-\frac{1}{2}\right)}{\Gamma^{\4}(\lambda)}.
\end{eqnarray}

By identifying the symplectic invariants with the determinant of $\sigma$, $\Sigma = (\alpha_s^{\2} -\gamma_s^{\2})(\alpha_q^{\2} -\gamma_q^{\2})$ and the diagonal and off-diagonal sub-determinants, $\Sigma_{\alpha} = \alpha_s\alpha_q$ and $\Sigma_{\gamma} = \gamma_s\gamma_q$, quantumness and separability -- according to the Peres-Horodecki (PH) criterium \cite{PH} for continuous variables \cite{Simon,Adesso} -- can be quantified through the computation of symplectic eigenvalues written in terms of $\alpha_{s,q}$ and $\gamma_{s,q}$ as
\begin{eqnarray}
d_{\pm} &=&
\sqrt{\frac{\Delta\pm\sqrt{\Delta^{\2}-\Sigma^{\2}}}{2}} = \sqrt{(\alpha_s \pm\gamma_s)(\alpha_q \pm\gamma_q)},\\
\tilde{d}_{\pm} &=& \sqrt{\frac{\tilde{\Delta}\pm\sqrt{\tilde{\Delta}^{\2}-\Sigma^{\2}}}{2}} = \sqrt{(\alpha_s \mp\gamma_s)(\alpha_q \pm\gamma_q)},
\end{eqnarray}
with $\Delta = 2(\Sigma_{\alpha}+\Sigma_{\gamma})$ and $\tilde{\Delta} = 2(\Sigma_{\alpha}-\Sigma_{\gamma})$.
The result for $\tilde{\Delta}$ is associated with a mirror reflection of the momentum coordinate of the subsystem $B$. It corresponds to the extension of the positive partial transposition (PPT) criterium for continuous variables \cite{Simon}.
The quantumness associated to the Robertson-Schr\"odinger uncertainty principle is consistent with $d_-\geq 1/2$, and the PH separability is consistent with $\tilde{d}_-\geq 1/2$. 

\begin{figure}[h]
\includegraphics[scale=0.45]{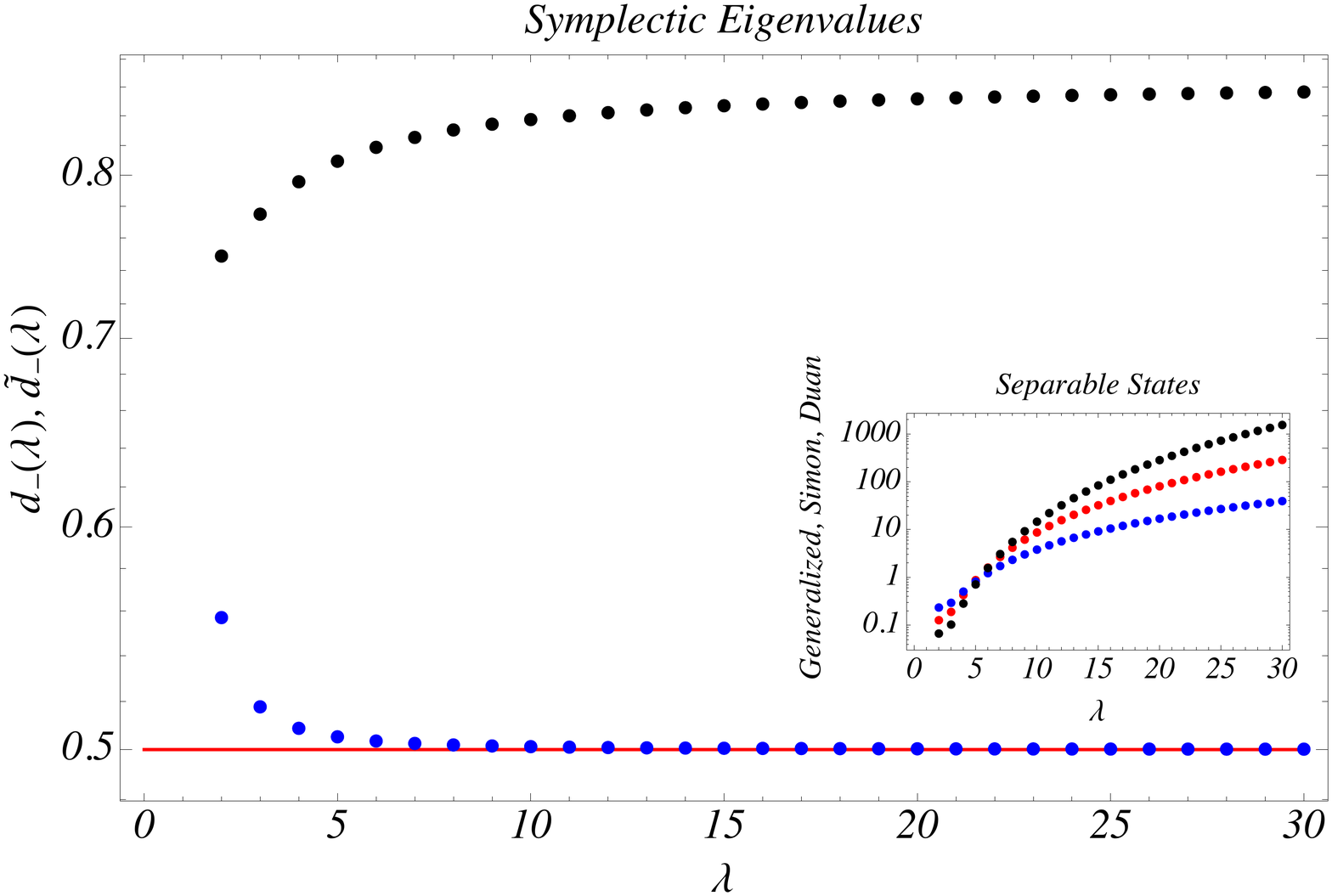}
\renewcommand{\baselinestretch}{.85}
\caption{(Color online) (Main Panel) Results for symplectic eigenvalues, $d_-$ (blue circles) and $\tilde{d}_-$ (black circles), as function of $\lambda$. For $\tilde{d}_- \geq 1/2$ the (Simon)-Peres-Horodecki criterium for separability is satisfied (given the solid red line as a boundary from below for $\tilde{d}_- = 1/2$). (Small Rectangle) The result is consistent with the generalized criterium which sets $Det[M_{\tilde{g}}(\rho))^\Gamma] \geq 0$ (black circles) for separable states, which is also consistent with the complementary criteria from Simon (modified) \cite{Simon2} (blue circles) and Duan \cite{Duan} (red circles).}
\label{Criterion}
\end{figure}

Fig.~\ref{Criterion} shows the results for lower symplectic eigenvalues $d_{-}$ and $\tilde{d}_{-}$, as function of $\lambda$, from which no entanglement profile according to the PH criterium is observed. Given that $d_{-} = 1/2$ saturates Robertson-Schr\"odinger uncertainty relation for a Gaussian state, in the context of a Gaussian approach, for $d_{-} > 1$, the quantumness is preserved.
Through a similar analysis, by noticing from Fig.~\ref{Criterion} that the symplectic eigenvalues, $\tilde{d}_-$, approach from above the red line for $\tilde{d}_- = 1/2$, the (Simon)-Peres-Horodecki criterium for separability is satisfied for all values of $\lambda$.
However, for non-Gaussian states, genuine entanglement may be revealed only through the application of criteria involving higher-order moments, which corresponds to an extension of the above-mentioned PPT criterium \cite{Miranowicz}.

By defining the adimensional operators $a$ and $a^\dagger$ for the subsystem $A$, and $b$ and $b^\dagger$ for the subsystem $B$, as
\begin{eqnarray}
a &=& \frac{1}{\sqrt{2}} \left( s_{_A} + i q_{_A} \right), \quad a^\dagger = \frac{1}{\sqrt{2}} \left( s_{_A} - i q_{_A} \right), \\
b &=& \frac{1}{\sqrt{2}} \left( s_{_B} + i q_{_B} \right), \quad b^\dagger = \frac{1}{\sqrt{2}} \left( s_{_B} - i q_{_B} \right), \label{eq:oper-b}
\end{eqnarray}
an auxiliary matrix of moments, $M_{g}(\rho) = [M_{ij}] = [\langle g_i^\dagger g_j \rangle]$ forms the basis for the criterium \cite{Miranowicz} which sets that, for $\rho^\Gamma$ denoting the partial transposition of the state $\rho$ with respect to the subsystem $B$, a bipartite state $\rho$ is entangled if and only if there exists $g$ such that the determinant of $M_g(\rho^\Gamma)$ is negative.
In this case, if the class $g$ of operators has a tensor product structure, $\tilde{g}=g^A \otimes g^B$, then $M_{\tilde{g}}(\rho^\Gamma) = (M_{\tilde{g}}(\rho))^\Gamma$.

For $\tilde{g}=(1, a) \otimes (1, b) = (1, a, b, ab)$, the corresponding matrix of moments becomes
\begin{eqnarray} \label{matrix}
M_{\tilde{g}}(\rho) &=& 
\left(
\begin{array}{cccc}
1 & \langle a \rangle & \langle b \rangle & \langle ab \rangle \\
\langle a^\dagger \rangle & \langle a^\dagger a \rangle & \langle a^\dagger b \rangle & \langle a^\dagger ab \rangle \\
\langle b^\dagger \rangle & \langle ab^\dagger \rangle & \langle b^\dagger b \rangle & \langle a b^\dagger b \rangle \\
\langle a^\dagger b^\dagger \rangle & \langle a^\dagger a b^\dagger \rangle & \langle a^\dagger b^\dagger b \rangle & \langle a^\dagger a b^\dagger b \rangle\end{array}
\right)=
\left(
\begin{array}{cccc}
1 & 0 & 0 & \frac{\gamma_s+\gamma_q}{4} \\
0& \frac{\alpha_s+\alpha_q}{4}&\frac{\gamma_s-\gamma_q}{4}&0\\
0& \frac{\gamma_s-\gamma_q}{4}&\frac{\alpha_s+\alpha_q}{4}&0\\
\frac{\gamma_s+\gamma_q}{4} & 0& 0 &\frac{(\alpha_s+\alpha_q)^{\2}}{16} 
\end{array}
\right),
\end{eqnarray}
which has been computed in terms the results from Eqs.~\eqref{averaged2}-\eqref{averaged} through the relations from \eqref{eq:oper-b}, which gives the entries of $(M_{\tilde{g}}(\rho))^\Gamma$ in terms of the statistical moments of the momentum and position variables.

The results for the determinant of $M_{\tilde{g}}(\rho))^\Gamma$ are also shown in Fig.~\ref{Criterion}, and they are complemented by two additional separability criteria, respectively, a modification to the Simon criterium \cite{Simon2} and the Duan criterium \cite{Duan}. Both involve the computation of the determinant of submatrices from $M_{\tilde{g}}(\rho))^\Gamma$, 
\begin{equation} 
M^{Mod}_{(\rho)} = 
\left(
\begin{array}{ccc}
1 & \langle b \rangle & \langle a b^{\dagger} \rangle\\
\langle b^\dagger \rangle & \langle b^\dagger b \rangle & \langle abb^{\dagger} \rangle \\
\langle a^{\dagger}b \rangle & \langle a^{\dagger}b^{\dagger}b \rangle & \langle a^\dagger a b^\dagger b \rangle \end{array}
\right)
\,\,\mbox{and}\,\,
M^{Duan}_{(\rho)} = 
\left(
\begin{array}{ccc}
1 & \langle a \rangle & \langle b^{\dagger} \rangle\\
\langle a^\dagger \rangle & \langle a^\dagger a \rangle & \langle a^\dagger b^{\dagger} \rangle \\
\langle b \rangle & \langle ab \rangle & \langle b^\dagger b \rangle \end{array}
\right),
\end{equation}
which again reproduce qualitatively the same results which imply the separability of the bipartite quantum state for all quantum numbers $\lambda$.
For completeness, it is worth mentioning that the spacial splitting of systems $A$ and $B$ does not affect the separability status of the above results since it leads to increasing values of the symplectic determinants and, of course, the suppression of the diagonal contributions from the pure state from Eq.~(\ref{wigner}) leads to mixed states which naturally exhibit entanglement profiles which can be theoretically manipulated according to the analytical expressions obtained in this paper.

To summarize, from the qualitative point of view, all the above results indeed do not increment any continuous variable entanglement effect with respect to which is observed for harmonic systems (i.e. when ground and first excited states from (\ref{state}) are replaced by the harmonic oscillator correspondent ones).
In fact, turning back to the overall analysis of what was proposed along this manuscript, the discussion of anharmonic aspects related to nonlinearity of quantum oscillators continuously provides fruitful discussions from both experimental applicability and theoretical conception point of views.
For instance, non-Gaussianity and fidelity-based measures based on the properties of the ground states rather than on the form of the potential have been theoretically tested as quantificators of nonlinearity for quantum oscillators \cite{ParisGeral} which include the PT one.
Through this framework, the nonlinearity features can be captured independently on the specific nature of the potential. 
In spite of addressing to the same issue which circumvents the connection between nonlinearities and anharmonic oscillations, differently from the above mentioned framework, the Wigner flow approach here discussed does not allow for the factorization of the potential contribution in the quantitative analysis. Besides driving the quantum perturbations, the contributions of
the potential and Wigner function coupled derivatives in the composition of the Wigner currents have been demonstrated to be relevant in the quantification of the fluxes of quantum information for anharmonic systems \cite{Our01,Our02}, an aspect which
certainly deserves more careful investigations.

\section{Conclusions}

Quantifiers of non-classicality, non-Gaussianity and overall quantum correlations are essential ingredients for quantum enhanced technologies and experiments involving macroscopic quantum coherence.
Considering exact solutions of an anharmonic system -- the hyperbolic PT quantum system -- in the phase-space, the quantum features distorting the corresponding classical portrait were identified for a ground and first excited state two-level system superposition.
The non-Liouvillian behavior was identified through the phase-space pattern of quantum fluctuations, and the non-Gaussian profile was quantified by measures of {\em kurtosis} and {\em negative entropy}.
Through the Wigner formalism in the phase-space, a monotonic relation between the entropic nonlinearity and non-classicality was applied to classifying ground and first excited states, and corresponding quantum superpositions.
Finally, a phase-space description of a bipartite quantum system of two particles in the PT potential revealed their separability properties under Gaussian as well as non-Gaussian approaches.
The experimental feasibility of PT potentials in double layer graphene \cite{Park15} and other experimental platforms \cite{Mick11,Al11} are indeed relevant in testing the limits of Gaussian and non-classical effects \cite{Bohem}, and verifying their inherent interplay with supersymmetric non-linear quantum mechanics \cite{Pana15} and entropic information scenarios \cite{Correa:2016pgr,Bernardini:2016hvx,Bernardini:2016qit}. In this context, the analytical results here obtained can be worked out into more involved configurations, for instance, those which could include three or more excited states, so as to engender some manipulable platform to investigate more than two-qubit quantum systems of continuous variables.

\vspace{.5 cm}
{\em Acknowledgments -- The work of AEB is supported by the Brazilian Agencies FAPESP (Grant No. 2018/03960-9) and CNPq (Grant No. 300831/2016-1). RdR is grateful to CNPq (Grant No. 303293/2015-2) and to FAPESP (Grant No.~2017/18897-8), for partial financial support.}


\begin{thebibliography}{99}
\bibitem{PT}
G. P\"oschl and E. Teller, Z. Phys. {\bf 83}, 143 (1933).
\bibitem{BookA}
G. B. Whitham, {\em Linear and Nonlinear Waves}, (Wiley, New York, 1974).
\bibitem{Bas01}
D. Bazeia, L. Losano and J. M. C. Malbouisson, Phys. Rev. D {\bf 66}, 101701 (2002).
\bibitem{Bazeia}
D. Bazeia, J. Menezes and R. Menezes, Phys. Rev. Lett. {\bf 91}, 241601 (2003).
\bibitem{AlexRoldao}
A. E. Bernardini and R. da Rocha, AHEP {\bf 2013}, 304980 (2013).
\bibitem{AlexRoldao2}
A. E. Bernardini and R. da Rocha, Phys. Lett. A {\bf 380}, 2279 (2016).
\bibitem{Pana15}
P. G. Kevrekidis, J. Cuevas-Maraver, A. Saxena, F. Cooper and A. Khare, Phys. Rev. E {\bf 92}, 042901 (2015).
\bibitem{Smith}
C. J. Pethick and H. Smith, {\em Bose-Einstein Condensation in Dilute Gases} (Cambridge University Press, Cambridge, 2001).
\bibitem{Ber12}
A. E. Bernardini and R. da Rocha, Phys. Lett. B {\bf 717}, 238 (2012).
\bibitem{Das}
A. Das, {\em Integrable Models} (World Scientific, Singapore, 1989).
\bibitem{Jatkar}
D. P. Jatkar, Nucl. Phys. B {\bf 395}, 167 (1993).
\bibitem{Witten}
E. Witten, Phys. Rev. D {\bf 44}, 314 (1991).
\bibitem{Gremm}
M. Gremm, Phys. Lett. B {\bf 478}, 434 (2000).
\bibitem{Barb08}
N. Barbosa-Cendejas, A. Herrera-Aguilar, M. A. Reyes Santos and C. Schubert, Phys. Rev. D {\bf 77}, 126013 (2008).
\bibitem{Bertolami}
A. E. Bernardini and O. Bertolami, Phys. Lett. B {\bf 726}, 512 (2013).
\bibitem{PLAetc}
J. Radovanovic, V. Milanovic, Z. Ikonic and D. Indjin, Phys. Lett. A {\bf 269}, 179 (2000).
\bibitem{Rasa12}
E. Rasanen, T. Blasi, M. F. Borunda and E. J. Heller, Phys. Rev. B {\bf 86}, 205308 (2012).
\bibitem{Mick11}
T. Micklitz and A. Levchenko, Phys. Rev. Lett. {\bf 106}, 196402 (2011).
\bibitem{Park15}
C.-S. Park, Phys. Rev. B {\bf 92}, 165422 (2015).
\bibitem{Sanc19}
O. de los Santos S\'anchez, J. Phys. A: Math. Theor. {\bf 51}, 305303 (2018).
\bibitem{Yildirim}
H. Yildirim and M. Tomak,
Phys. Rev. B {\bf 72}, 115340 (2005); J. App. Phys. {\bf 99}, 093103 (2006).
\bibitem{Al11}
S. M. Al-Marzoug, S. M. Al-Amoudi, U. Al Khawaja, H. Bahlouli, and B. B. Baizakov, Phys. Rev. E {\bf 83}, 026603 (2011).
\bibitem{Wigner}
E. Wigner, Phys. Rev. {\bf 40}, 749 (1932).
\bibitem{Curtright}
T. Curtright, D. Fairlie and C. Zachos, Phys. Rev. D {\bf 52}, 025002 (1998).
\bibitem{Bund}
G. W. Bund and M. C. Tijero, Phys. Rev. A {\bf 61}, 052114 (2000).
\bibitem{Case}
W. B. Case, Am. J. Phys. {\bf 76}, 937 (2008). 
\bibitem{Steuernagel3}
O. Steuernagel, D. Kakofengitis and G. Ritter, Phys. Rev. Lett. {\bf 110}, 030401 (2013).
\bibitem{Liouvillian}
D. Kakofengitis, M. Oliva and O. Steuernagel, Phys. Rev. A {\bf 95}, 022127 (2017).
\bibitem{Our01}
A. E. Bernardini and O. Bertolami, Europhysics Letters (EPL) {\bf 120}, 20002 (2017).
\bibitem{Our02}
A. E. Bernardini, P. Leal and O. Bertolami, JCAP {\bf 02}, 025 (2018).
\bibitem{Ferraro11}
A. Ferraro and M. G. A. Paris, Phys. Rev. Lett. {\bf 108}, 260403 (2012).
\bibitem{Donoso12}
A. Donoso and C. C. Martens, Phys. Rev. Lett. {\bf 87}, 223202 (2001).
\bibitem{Tunnel}
N. L. Balazs and A. Voros, Annals Phys. {\bf 1}, 123 (1990).
\bibitem{Ballentine}
L. E. Ballentine, {\em Quantum Mechanics: a Modern Development} (World Scientific, Singapore, 1998).
\bibitem{Alba18}
F. Albarelli, M. G. Genoni, M. G. A. Paris and A. Ferraro, {\em Resource theory of quantum non-Gaussianity and Wigner negativity}, arXiv:1804.05763 [quant-ph].
\bibitem{Park18}
J. Park, J. Lee, K. Baek, S-W Ji, and H. Nha, {\em Faithful measure of Quantum non-Gaussianity via quantum relative entropy}, arXiv:1809.02999 [quant-ph].
\bibitem{Park17}
J. Park, J. Lee, S-W Ji, and H. Nha, Phys. Rev. A {\bf 96}, 052324 (2017).
\bibitem{Gradshteyn}
I. S. Gradshteyn and I. Ryzhik, {\it Tables of Integrals, Series and Products} (Academic Press, New York, 1994).
\bibitem{Domcke}
{\em Driven Quantum Systems}, edited by W. Domcke, P. H\"anggi, and D. Tannor, Chem. Phys. {\bf 217}, 117 (1997).
\bibitem{Alba16}
F. Albarelli, A. Ferraro, M. Paternostro and M. G. A. Paris, Phys. Rev. A {\bf 93}, 032112 (2016).
\bibitem{Genoni00}
C. Hughes, M. G. Genoni, T. Tufarelli, M. G. A. Paris and M. S. Kim, Phys. Rev. A {\bf 90}, 013810 (2014).
\bibitem{Genoni01}
M. G. Genoni, M. G. A. Paris and K. Banaszek, Phys. Rev. A {\bf 76}, 042327 (2007).
\bibitem{Genoni02}
M. G. Genoni, M. G. A. Paris and K. Banaszek, Phys. Rev. A {\bf 78}, 060303 (2008).
\bibitem{Olivares}
A. Ferraro, S. Olivares, and M. G. A. Paris, {\em Gaussian states in continuous variable quantum information} (Bibliopolis, Napoli, 2005).
\bibitem{Holevo}
A. S. Holevo, M. Sohma, and O. Hirota, Phys. Rev. A {\bf 59}, 1820 (1999).
\bibitem{Serafini}
A. Serafini, F. Illuminati, and S. De Siena, J. Phys. B {\bf 37}, L21 (2004).
\bibitem{Bertolami2017}
R. Alves, C. Bastos and O. Bertolami, {\em Entangled States and the Gravitational Quantum Well} [arXiv:1607.08155 [gr-qc]].
\bibitem{PH}
A. Peres, Phys. Rev. Lett. {\bf 77}, 1413 (1996); P. Horodecki, Phys. Lett. A {\bf 232}, 333 (1997).
\bibitem{Simon}
R. Simon, E. C. G. Sudarshan, and N. Mukunda, Phys. Rev. A {\bf 36}, 3868 (1987).
\bibitem{Adesso}
G. Adesso, A. Serafini, and F. Illuminati, Phys. Rev. Lett. {\bf 92}, 087901 (2004).
\bibitem{Miranowicz}
A. Miranowicz, M. Piani, P. Horodecki, and R. Horodecki, Phys. Rev. A {\bf 80}, 052303 (2009).
\bibitem{Simon2}
R. Simon, Phys. Rev. Lett. {\bf 84}, 2726 (2000).
\bibitem{Duan}
L.-M. Duan, G. Giedke, J. I. Cirac, and P. Zoller, Phys. Rev. Lett. {\bf 84}, 2722 (2000).
\bibitem{ParisGeral}
M. G. A. Paris, M. G. Genoni, N. Shammah, and B. Teklu, Phys. Rev. A {\bf 90}, 012104 (2014).
\bibitem{Bohem}
S. Dey and A. Fring, Phys. Rev. A {\bf 88}, 022116 (2013).
\bibitem{Bernardini:2016hvx}
A. E. Bernardini and R. da Rocha, Phys. Lett. B {\bf 762}, 107 (2016).
\bibitem{Correa:2016pgr}
R. A. C. Correa, D. M. Dantas, C. A. S. Almeida and R. da Rocha, Phys. Lett. B {\bf 755}, 358 (2016).
\bibitem{Bernardini:2016qit}
A. E. Bernardini, N. R. F. Braga and R. da Rocha, Phys. Lett. B {\bf 765}, 81 (2017).

\end{thebibliography}
 \end{document}